\documentstyle[onecolumn]{mn}

\newcounter{parentequation}
\def\beglet{
  \addtocounter{equation}{1}%
  \setcounter{parentequation}{\value{equation}}%
  \setcounter{equation}{0}%
  \def\theequation{\arabic{parentequation}\alph{equation}}%
  \ignorespaces
}
\def\endlet{
  \setcounter{equation}{\value{parentequation}}%
  \def\theequation{\arabic{equation}}%
}
\def\pmb#1{\setbox0=\hbox{#1}%
    \kern-.025em\copy0\kern-\wd0
    \kern.05em\copy0\kern-\wd0
    \kern-.025em\raise.0433em\box0}

\def\ltsima{$\; \buildrel < \over \sim \;$}
\def\gtsima{$\; \buildrel > \over \sim \;$}
\def\simlt{\lower.5ex\hbox{\ltsima}}
\def\simgt{\lower.5ex\hbox{\gtsima}}
\def\p2Y{\;_2Y}
\def\m2Y{\;_{-2}Y}

\def\etal{{\it et al.}\rm}
\def\etals{{\it et al. }\rm}
\def\mk2{\mu {\rm K}^2}

\begin{document}

\title[Correlations of the CMB]
{Large-Angle Correlations in the Cosmic Microwave Background}

\author[Efstathiou, Ma and Hanson]{George Efstathiou,  Yin-Zhe Ma and 
Duncan Hanson \\
 Kavli Institute for Cosmology Cambridge and 
Institute of Astronomy, Madingley Road, Cambridge, CB3 OHA.}

\maketitle

\begin{abstract}
  It has been argued recently by Copi \etals (2009) that the lack of
  large angular correlations of the CMB
  temperature field provides strong evidence against the standard,
  statistically isotropic, inflationary $\Lambda$CDM cosmology. We
  compare various estimators of the temperature correlation function
  showing how they depend on assumptions of statistical isotropy and
  how they perform on the WMAP 5 year ILC maps with and without a sky
  cut. We show that the low multipole harmonics that determine the
  large-scale features of the temperature correlation function can be 
  reconstructed accurately, independent of any assumptions concerning
  statistical isotropy, from the data that lie outside the sky
  cuts. The temperature correlation functions computed from our
  reconstructions are in good agreement with those computed from the
  whole sky. A Bayesian analysis of the large-scale correlations is
  presented which shows that the data cannot exclude the standard
  $\Lambda$CDM model. We discuss the differences between our
  conclusions and those of Copi \etal.

\vskip 0.1 truein

\noindent
{\bf Key words}: 
Methods: data analysis, statistical; Cosmology: cosmic microwave background,
large-scale structure of Universe

\vskip 0.3 truein

\end{abstract}

\section{Introduction}

Following the discovery by the COBE team of temperature
anisotropies in the cosmic microwave background (CMB) radiation (Smoot
\etals 1992; Wright \etals 1992), Hinshaw \etals (1996) noticed that
the temperature angular correlation function,  $C(\theta)$,  measured
from the  COBE maps was close to zero on large angular scales. This
result attracted little attention until the publication of the first
year results from WMAP (Bennett \etals 2003; Spergel \etals 2003, hereafter S03). The
results from WMAP confirmed the lack of large-scale angular
correlations in the temperature maps and led S03 to
introduce the statistic 
\begin{equation}
  S_{1/2} = \int_{-1}^{1/2} [C(\theta)]^2 d\cos \theta. \label{C12}
\end{equation}
The form of the statistic and the upper cut-off, $\mu=\cos \theta = 1/2$, were
chosen {\it a posteriori} by S03 `in response' to the observed shape
of the temperature correlation function computed using a particular
estimator and sky cut (as described in further detail in Section
2). To assess the statistical significance of the lack of large-scale
power, S03 computed a `p-value', {\it i.e.} the fraction of models 
 in their Monte Carlo Markov chains  which had a value of
$S_{1/2}^{\rm model} < S_{1/2}^{\rm data}$, using the same estimator
and sky cut that they applied to the data. For their standard
six-parameter inflationary $\Lambda$CDM cosmology, they found a
p-value of $0.15\%$, suggesting a significant discrepancy between the
model and the data.

This problem was revisited by Efstathiou (2004a, herafter E04). The
main focus of the E04 paper was to improve on the pseudo-harmonic
power spectrum analysis used by the WMAP team (Hinshaw \etals 2003)
 by using quadratic maximum likelihood (QML)
estimates of the power spectrum, with particular emphasis on the
statistical significance of the low amplitude of the quadrupole
anisotropy. As an aside, E04 computed angular correlation functions
from the QML power spectrum estimates and showed that they were
insensitive to the presence of a sky cut. Similarly the $S_{1/2}$
statistic computed from these correlation functions was found to be
insensitive to the size of the sky cut giving p-values of a few
percent. E04 concluded that the correlation function and $S_{1/2}$
statistic offered no compelling evidence against the concordance
inflationary $\Lambda$CDM model. E04 did not explore in any detail the
low p-value for the $S_{1/2}$ statistic reported by S03
(2003), but commented that it was probably simply an `unfortunate'
consequence of the particular choice of statistic, estimator and sky
cut chosen by these authors (in other words, a result of various
{\it a posteriori} choices).

The CMB temperature correlation function and $S$ statistic have been
reanalysed in two recent papers (Copi \etals 2007; Copi \etals 2009,
heafter CHSS09).  The arguments in the two papers are quite similar,
and so for the most part we will refer to the later paper (since, as
in this work, it analyses the 5 year WMAP temperature data, Hinshaw
\etals 2009). The Copi \etals papers are largely motivated by evidence
for a violation of statistical isotropy in the WMAP temperature maps,
in particular evidence of alignments amongst the low order CMB
multipoles ({\it e.g.} Tegmark, de Oliveira-Costa and Hamilton 2003;
Schwarz \etals 2004; Land and Magueijo 2005a, b), although the
statistical significance of these alignments has been questioned (de
Oliveira-Costa \etals 2004; Francis \& Peacock 2009).  Copi \etals
make the valid point that statistical isotropy is often implicitly
assumed in defining what is meant by the term `correlation function'
and in defining estimators. They argue further that different
estimators contain different information.  They then focus on
pixel-based estimates of the correlation function applied to the WMAP
data, including a sky cut, and find p-values for the $S_{1/2}$
statistic of $\sim 0.025$--$0.04\%$, depending on the choice of CMB
map and sky cut. If no sky cut is applied, they find p-values of $\sim
5\%$ (similar to the p-values reported in E04). CHSS09 comment that
the full-sky results are apparently inconsistent with the cut-sky
analysis suggesting a violation of statistical isotropy.

Any analysis which claims to strongly rule out the simple inflationary
$\Lambda$CDM model deserves careful scrutiny, since a confirmed
discordance would have profound consequences for our understanding of
the early Universe. The purpose of this paper is to investigate
carefully the analysis presented in CHSS09. In Section 2 we discuss
estimators of the correlation function and relate the pixel-based
estimator used by CHSS09 to the pseudo-power spectrum computed on a
cut sky. In Section 3, we explicitly reconstruct the individual low
order multipole coefficients $a_{\ell m}$ from cut-sky maps using a
technique first applied by de Oliveira-Costa and Tegmark (2006). This
allows us to test the sensitivity of the large-angle correlation
function to the presence of a sky cut, independent of any assumptions
concerning statistical isotropy or Gaussianity. The results of this
analysis are compared with the QML estimates of the correlation
function used in E04.  Section 4 describes a Bayesian analysis of the
$S_{1/2}$ statistic and contrasts it with the frequentist analysis
applied by S03 and CHSS09.  Our conclusions are summarized in Section
5.

\section{Estimators of the Correlation Function}

If we assume statistical isotropy, the ensemble average of the
temperature angular correlation function (ACF) measured over the whole
sky $\langle C(\theta)\rangle$ is related to the ensemble average of the angular power
spectrum $\langle C_\ell \rangle$ by the well known relation
\begin{equation}
\langle C(\theta) \rangle = {1 \over 4 \pi} \sum_\ell(2 \ell + 1)
\langle C_\ell \rangle P_\ell (\cos \theta). \label{C0}
\end{equation}
However, we have only one realization of the sky and we may further
choose to impose a sky cut to reduce possible contamination
from regions of high Galactic emission.  If we relax the assumptions
of statistical isotropy and complete sky coverage,  there is no
unique definition or estimator of the ACF. One can write down a number of 
  estimators that, given certain assumptions concerning
the underlying statistics of the fluctuations, may average to the
ensemble mean when applied to data on an incomplete sky.

CHSS09 use a direct pixel based correlation function\footnote{
More generally, one can define a bi-polar correlation function 
 $C(\pmb{$\theta$}_i, \pmb{$\theta$}_j
$), which can be used
as a test of statistical isotropy (see {\it e.g.} Basak, Hajian
and Souradeep (2006).}
on the cut sky
\begin{equation}
 C^{\rm pix}(\theta) = \langle x_i x_j \rangle, \label{C1}
\end{equation}
where $x_i$ denotes the  temperature value in pixel $i$ and the
angular brackets  denote an average over all pixel
pairs outside the sky cut with an angular separation that lies within
a small interval of $\theta$.

If the underlying temperature field is statistically isotropic,
equation (\ref{C1}) provides an unbiased estimate of the correlation
function, i.e. the average over a large number of independent
realizations is unbiased, irrespective of the sky cut. However, if the
fluctuations are statistically isotropic and Gaussian, (\ref{C1}) is
not an optimal estimator of $\langle C(\theta) \rangle$. To see this,
expand the temperature field in spherical harmonics
\begin{equation}
x_i = \sum_{\ell m} a_{\ell m}Y_{\ell m}(\pmb{$\theta$}_i ), \qquad \langle \vert a_{\ell m} \vert^2 \rangle = \langle C_\ell \rangle. \label{C2}
\end{equation}
Then, from the rotation properties of the spherical harmonics, it is
straightfoward to prove
\begin{equation}
  C^{\rm pix}(\theta_{ij}) = \langle x_i x_j \rangle = {\sum_\ell (2 \ell + 1) \tilde C^P_\ell P_\ell(\cos \theta_{ij})
    \over \sum_\ell (2 \ell + 1) \tilde W_\ell P_\ell (\cos \theta_{ij})}, \label{C3}
\end{equation}
where $\tilde C^P_\ell$ is the pseudo-power spectrum (PCL) estimate on the cut 
sky:
\begin{equation}
\tilde C^P_\ell = {1 \over (2 \ell + 1)} \sum_m \vert \tilde a_{\ell m} \vert^2,
\quad \tilde a_{\ell m} = \sum_i x_i w_i Y^*_{\ell m}(\pmb{$\theta$}_i )\Omega_i, \label{C4}
\end{equation}
where $w_i$ is a window function that is zero or unity depending on
whether a pixel (of area $\Omega_i$) lies inside or outside the 
sky cut.  The function $\tilde W_\ell$ in (\ref{C3}) is the pseudo-power spectrum
of the window function $w_i$:
\begin{equation}
\tilde W_\ell = {1 \over (2 \ell + 1)} \sum_m \vert \tilde w_{\ell m} \vert^2,
\quad \tilde w_{\ell m} = \sum_i  w_i 
\Omega_iY^*_{\ell m}(\pmb{$\theta$}_i ). \label{C4a}
\end{equation}
The coefficients $\tilde a_{\ell m}$ are related to the
coefficients $a_{\ell m}$ on the uncut sky by the coupling matrix
${\bf K}$,
\begin{equation}
  \tilde a_{\ell m} = \sum_{\ell^\prime m^\prime} a_{\ell^\prime m^\prime} K _{\ell m \ell^\prime m^\prime}, \label{C5}
\end{equation}
where 
\begin{equation}
  K_{\ell_1 m_1 \ell_2 m_2} = \sum_i w_i \Omega_i Y^*_{\ell_1 m_1} (\pmb{$\theta$}_i) Y_{\ell_2 m_2} (\pmb{$\theta$}_i). \label{C6}
\end{equation}
The relation (\ref{C3}) is an identity and does not depend on the
assumption of statistical isotropy.

Under the assumption of statistical isotropy and Gaussianity, it is
straightfoward to calculate the covariance matrix of the estimator
(\ref{C1})
\begin{equation}
\langle \Delta \tilde C(\theta_i) \Delta \tilde C(\theta_j) \rangle =
\left ( \sum_{\ell_1 \ell_2}  (2 \ell_1 + 1) (2 \ell_2 + 1) \langle \tilde \Delta C^P_{\ell_1}
\tilde \Delta C^P_{\ell_2} \rangle P_{\ell_1} (\cos \theta_{ij}) P_{\ell_2} (\cos \theta_{ij}) \right )
/\left ( \sum_{\ell} (2 \ell+1) \tilde W_{\ell} P_{\ell} (\cos \theta_{ij}) \right)^2 \label{C7}
\end{equation}
where
\begin{equation}
\langle \Delta \tilde C^P_{\ell} \Delta \tilde C^P_{\ell^\prime} \rangle =
{2 \over (2 \ell + 1) (2 \ell^\prime + 1)} \sum_{m m^\prime} \sum_{\ell_1 m_1} \sum_{\ell_2 m_2} C_{\ell_1} C_{\ell_2} K_{\ell m \ell_1 m_1} K^*_{\ell^\prime m^\prime \ell_1 m_1}
K^*_{\ell m \ell_2 m_2} K_{\ell^\prime m^\prime \ell_2 m_2} \label{C8}
\end{equation}

\begin{figure*}
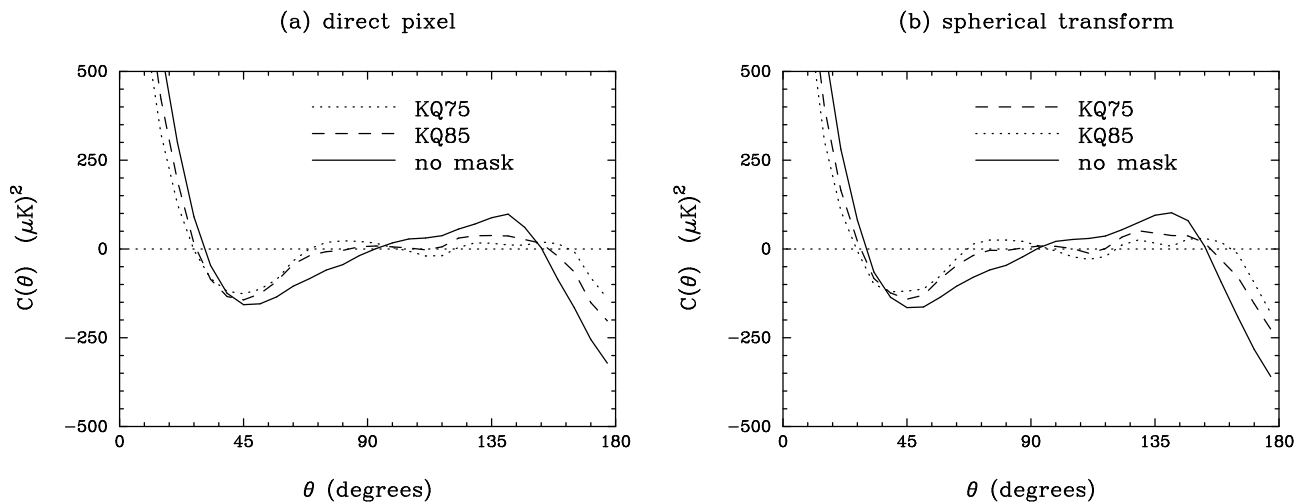

\vskip 3.0 truein

\includegraphics{pgctheta.ps}
\includegraphics{pgctheta_pcl.ps}

\caption {The figure to the left shows the correlation functions
  computed using the pixel estimator (\ref{C1}) applied to the 5 year WMAP
  ILC map (degraded as described in the text) over the full sky and
  with the WMAP KQ85 and KQ75 masks imposed.  The figure to the right
  shows the correlation functions computed from the pseudo-power
  spectra (\ref{C3}).}

\label{figure1}
\end{figure*}

Figure 1(a) shows the direct pixel based estimator (\ref{C1}) applied
to the 5 year WMAP ILC map after smoothing with a Gaussian filter of
$10^\circ$ FWHM and repixelising at a Healpix (Gorski \etal, 2005)
resolution NSIDE=16. The results of Figure 1 are consistent with those
of CHSS09. With the WMAP KQ85 and KQ75 masks\footnote{ We used
  degraded resolution (NSIDE=16) versions of these masks.  The
  degraded resolution KQ75 mask is plotted in Figure 4.} applied
(retaining about $82\%$ and $71\%$ of the sky respectively, Gold
\etals 2009), there is little power over the angular range
$60^\circ$-$160^\circ$. However, there is some non-zero correlation if
the pixel estimator is evaluated over the full sky. Figure 1(b) shows
the correlation functions determined from the pseudo-spectra
(\ref{C3}). This simply confirms the equivalence of the two estimators
(\ref{C1}) and (\ref{C3}) (apart from minor differences arising from
the finite angular bin widths). The covariance matrix for these 
estimators for the full sky is shown in Figure 2, using the $C_\ell$
for the six parameter $\Lambda$CDM model that provides the best fit to
the WMAP data (Komatsu \etals 2009).  The large angle ACF for a nearly
scale invariant temperature spectrum is dominated by a small number of
modes leading to large correlations between different angular scales.
The main effect of a KQ75-type sky cut on the covariance matrix is to
increase its overall amplitude.  The angular
structure of the covariance matrix is insensitive to the precise size
and shape of the sky cut.

The results plotted in Figure \ref{figure1} are very similar to those
presented by S03  and E04 using a slightly different estimator
\begin{equation}
  C^P(\theta) = {1 \over 4 \pi} \sum_\ell (2 \ell + 1) \hat C^P_{\ell}
  P_\ell (\cos \theta), \qquad \hat C^P_{\ell} =  M^{-1}_{\ell \ell^\prime} \tilde C^P_{\ell^\prime}, \label{C9}
\end{equation}
where the matrix $M$ is
\begin{equation}
M_{\ell \ell^\prime} = {1 \over (2 \ell + 1)} \sum_{ m m^\prime } \vert K_
{\ell m \ell^\prime m^\prime}
\vert^2. \label{C10}
\end{equation}
If we assume statistical isotropy, then both of the estimators (\ref
{C1}) and (\ref{C9}) are unbiased estimators of the ensemble mean ACF
and the power spectrum estimate $\hat C_\ell^P$ is an unbiased
estimate of the true power spectrum $\langle C_\ell\rangle$. (The
matrix ${\bf M}^{-1}$ `deconvolves' the PCL estimates $\tilde
C_\ell^P$ correcting the bias introduced by a sky cut, Hivon
\etals 2002).

One can see from (\ref{C3}) that the estimators (\ref{C1}) and
(\ref{C9}) are identical for the complete sky, since then
\begin{equation}
 w_{\ell m} = \left ( {2 \pi  } \right ) ^{1/2} \delta_{\ell 0} \delta_{m 0},
\quad M_{\ell \ell^\prime} = \delta_{\ell \ell^\prime}. \label{C11}
\end{equation}
The estimators (\ref{C1}) and (\ref{C9}) will, however, differ if a sky
cut is applied to the data. Nevertheless, since both estimators
involve sums over pseudo-spectral coefficients, and since they are
formally identical for the full sky, one would expect that they would
return similar estimates when applied to data incorporating moderate sky cuts
such as the KQ85 and KQ75 masks. 

This is, indeed, what we find. Figure 3 shows the estimator (\ref{C9})
applied to the WMAP 5 year ILC map. The results are almost identical
to those shown in Figure 1. The pixel estimator of the ACF (\ref{C1})
is therefore almost identical to the estimator of (\ref{C9}) and so,
as expected, it contains little new information (regardless of any
assumptions concerning statistical isotropy).

\begin{figure*}

\vskip 2.8 truein

\includegraphics{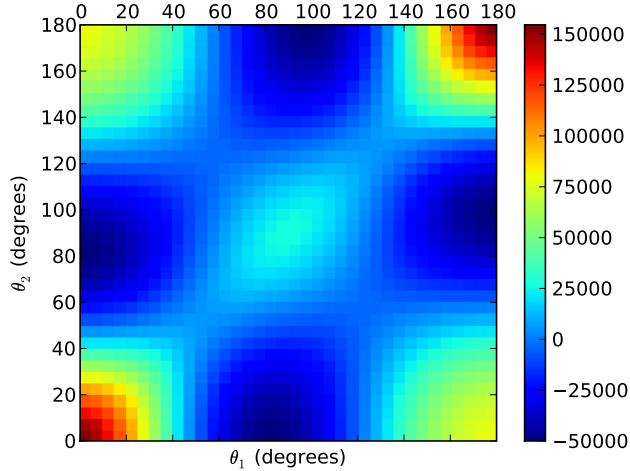}

\caption
{Covariance matrix for the pixel correlation estimator
(\ref{C1}) computed for full sky maps. The scale on the right is
in units of $(\mu {\rm K})^4$.}

\label{figure2}
\end{figure*}

\begin{figure*}
\vskip 3.0 truein

\includegraphics{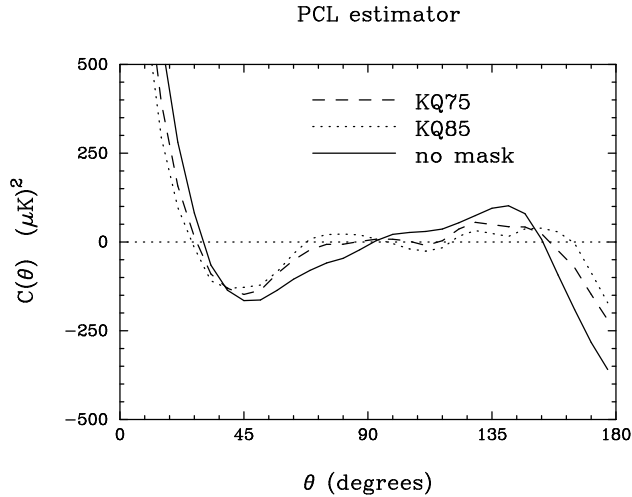}

\caption
{As in Figure 1 but using the pseudo-harmonic estimator of equation (\ref{C9}).}

\label{figure3}
\end{figure*}

The main differences between the various ACF estimates plotted in
Figures 1 and 3 come from application of the sky cuts.  With the
sky cuts applied, the ACF's are close to zero on angular scales
$\simgt 60^\circ$. This lack of power leads to particulary low values
for the $S_{1/2}$ statistic of about $1000$-$2000\; (\mu{\rm K})^4$,
as listed in Table 1. If no sky cut is applied, the value of the
$S_{1/2}$ statistic is substantially higher at around $8000 \;
(\mu{\rm K})^4$. It is worth noting that the values listed in Table 1
are very similar to the values obtained from the WMAP first year data
({\it cf} Table 5 in E04). The low multipole anisotropies that
contribute to the ACF at large angular scales have remained stable as
the data have improved. The low multipoles are signal dominated and
stable to improved gain corrections, foreground separation and small
perturbations to the Galactic mask.

\begin{table}

\centerline{\bf \ \ \  Table 1:  Values of the S$_{\bf 1/2}$ statistic for WMAP 5 year ILC maps$^1$}

\begin{center}

\begin{tabular}{ccccccccc} \hline \hline
\smallskip 
  & \multicolumn{8}{c}{ $S_{1/2}$ statistic in $(\mu{\rm K})^4$} \cr
sky cut       & Pixel ACF  & Pixel ACF  & PCL ACF & \multicolumn{4}{c} {Harmonic Reconstruction} & QML ACF \cr
      & (equ \ref{C1}) &  (equ \ref{C3}) & (equ \ref{C9}) & $\ell_{\rm max}=5$&
$\ell_{\rm max}=10$ & $\ell_{\rm max}=15$ & $\ell_{\rm max}=20$ & (equ \ref{C13}) \cr
\hline \hline
full sky  & $7373$ & $8532$  & $8532$ & $8170$ & $7777$ & $7649$ & $7606$ & $8532$ \cr
KQ85    & $1401$ &  $1781$ & $1901$ & $8250$ & $6953$ & $7612$ & $6383$ & $7234$\cr
KQ75  & $647$ &  $963$ & $1010$ & $7913$ & $6914$ & $8233$ & $5139$ &$5764$ \cr
\hline
\end{tabular}

\end{center}
\begin{center}
  $\;$$^1$ For maps degraded to Healpix resolution of NSIDE$=16$ and
  smoothed with a Gaussian of FWHM $10^\circ$.
\end{center}
\end{table}

As discussed in the Introduction, CHSS09 argue that the pixel
estimates of the ACF computed from the masked regions of the sky lead
to p-values with respect to the standard $\Lambda$CDM cosmology of
$\sim 0.1\%$ or less, suggesting a significant discrepancy between the
model and the data. However, the p-values computed for the unmasked
sky are much less significant ($\sim 5\%$). Various interpretations of
this result have been proposed:

\noindent
(i) The interpretation put forward by CHSS09 is that either
correlations have been introduced in reconstructing the full sky maps
from the observations, or that there are highly significant departures
from statistical isotropy that are correlated with the Galactic sky
cut leading to an ACF that is very close to zero for regions outside
the sky cut.

\noindent
(ii) The interpretation put forward by E04 is that the low p-values
are a consequence of using a sub-optimal estimator of the ACF on a cut
sky, combined with  a posteriori choices of the form of the
$S_{1/2}$ statistic.  On this interpretation, the low p-values found
by S03, Copi \etals (2007) and CHSS09, are of no physical
significance.

In support of point (ii), E04 used a quadratic maximum likelihood (QML) estimator
of the power spectrum, $\hat C^Q_\ell$, and computed the ACF
\begin{equation}
C^Q(\theta) = {1 \over 4 \pi} \sum_{\ell} (2 \ell +1 ) \hat C^Q_\ell P_\ell (\cos \theta). \label{C13}
\end{equation}
For Gaussian temperature maps, the QML estimates $\hat C^Q_\ell$ have
a significantly smaller variance than the PCL estimates $\hat
C^P_\ell$ if a sky cut is applied to the data\footnote{For noise-free
  data over the full sky, the QML and PCL estimators are
  identical.}. Hence the estimator $C^Q(\theta)$ will generally be
closer to the truth ({\it i.e.} closer to the ensemble mean $\langle
C(\theta) \rangle$) than the estimator (\ref{C9}). Applied to the
first year WMAP ILC map, E04 found that the ACF estimates derived from
(\ref{C13}) are insensitive to a sky cut and lead to p-values for the
$S$-statistic of $\sim 5 \%$, {\it i.e.} no strong evidence against
the concordance  $\Lambda$CDM model.

The reason that the QML estimator has significantly smaller `estimator
induced' variance than the PCL estimator is easy to understand (see
Efstathiou 2004b). For noise-free band limited data, it is possible to
reconstruct the low multipole coefficients $a_{\ell m}$ exactly from
data over an incomplete sky. This is, in effect, what the QML
estimator does, though it implicitly assumes statistical isotropy in
weighting the $a_{\ell m}$ coefficients to form the
power-spectrum. For low multipoles, the assumption of statistical
isotropy is unimportant, and for the noise-free data and sky cuts
relevant to WMAP, the low order multipole coefficients and
the power spectrum $C_\ell$ can be reconstructed almost exactly from
data on the incomplete sky. In this paper, we will extend the analysis
of E04 by explicitly reconstructing the low order coefficients
$a_{\ell m}$ over the entire sky. This analysis will confirm that the
ACF at large angular scales is insensitive to a sky cut and leads to
p-values of marginal significance.

The `estimator induced' variance of the pixel estimator of the ACF
(\ref{C1}) is also easy to understand intuitively.  (This problem has
been discussed extensively in the literature in the context of angular
clustering analysis of galaxy surveys: Groth and Peebles 1986; Landy
and Szalay 1993; Hamilton 1993; Maddox, Efstathiou and Sutherland
1996). The ACF is a pair-weighted statistic. Consider the analysis of
data on an incomplete sky. An overdensity, or underdensity, close to
the boundary of the sky cut will almost certainly continue as an
overdensity, or underdensity, across the  cut. If the pair count is
merely corrected by the missing area that lies within the cut region
of sky (as in the estimator \ref{C1}) overdense and underdense regions
close to the boundary will be underweighted. This causes no bias to the
estimator, but increases the sample variance. The analysis presented
in the next Section shows that it is possible reduce this sampling variance by
reconstructing the low
multipoles across a sky cut in a way that is numerically stable and free of
assumptions concerning statistical isotropy.

\section{Reconstructing low-order multipoles on a cut sky}

The aim of this Section is to reconstruct the large-scale features of
the temperature anisotropies over the whole sky using only the
incomplete data that lies outside a chosen sky cut. This can be done
in a number of ways, for example, by Weiner or `power equalization'
filtering (Bielwicz, G\'orski and Banday, 2004), Gibbs sampling
(Wandelt, Larson and Lakshminarayanan, 2004; Eriksen \etals 2004) or
by `harmonic inpainting' (Inoue, Cabella and Komatsu, 2008). Here we
apply a direct inversion method, which is insensitive to assumptions
concerning the statistical properties of the temperature field.

Let the vector ${\bf x}$ denote the temperature field on the
sky and let the vector ${\bf a}$ denote the spherical harmonic coefficients
$a_{\ell m}$. The vectors ${\bf x}$ and ${\bf a}$ are related by the spherical
transform ${\bf Y}$,
\begin{equation}
{\bf x} ={\bf Y} {\bf a} + {\bf n},  \label{M1}
\end{equation}
where ${\bf n}$ represents `noise' in the data.

Now consider the reconstruction ${\bf a^e}$
\begin{equation}
{\bf a^e} = ({\bf Y}^T {\bf A} {\bf Y})^{-1} {\bf Y}^T{\bf A} {\bf x}, \label{M2}
\end{equation}
for any arbitrary square matrix ${\bf A}$. The reconstruction is related to
the true coefficients ${\bf a}$ by
\begin{equation} {\bf a^e} = {\bf a} + ({\bf Y}^T {\bf A} {\bf
    Y})^{-1} {\bf Y}^T{\bf A} {\bf n}. \label{M3}
\end{equation}
If the data is noise free, (\ref{M2}) recovers the true vector ${\bf a}$
exactly. If, further, we choose ${\bf A}$ to be the identity matrix, then
\begin{equation}
{\bf a^e} = {\bf K}^{-1}{ \bf \tilde a}, \label{M4}
\end{equation}
where ${\bf K}$ is the coupling matrix (\ref{C6}). 

The reconstruction of (\ref{M4}) is closely related to the problem of
defining an orthonormal basis set of functions on the cut sky, which
has been studied extensively in the literature (see {\it e.g.}
G\'orski 1994; G\'orski \etals 1994; Mortlock, Challinor and Hobson,
2002). If the sky cut is relatively small, and the data are noise-free
and band-limited, the coupling matrix ${\bf K}$ will be non-singular
and can be inverted to yield the full-sky harmonics ${\bf a}$ exactly.
If the data are noise-free but not band-limited, the matrix ${\bf K}$
will become numerically singular on the incomplete sky as $\ell_{\rm
  max} \rightarrow \infty$. (As a rule-of-thumb the matrix will become
singular if $\ell_{\rm max}$ exceeds the inverse of the width of the
sky cut in radians.) This simply tells us that there are `ambiguous'
harmonic coefficients that are unconstrained by the data outside the
sky cut. For noise-free data that are not strictly band-limited, the
solution (\ref{M2}) trunctated to a finite value of $\ell_{max}$ will
amplify some of the high frequency signal which will appear as `noise'
within the sky cut in the reconstruction ${\bf x^e} = {\bf Y a^e}$.  The
amplitude of this `noise' can be reduced by an appropriate choice of
the matrix ${\bf A}$. If we assume that the signal and noise are
Gaussian, the optimal solution of (\ref{M1}) is the familiar `map-making' solution 
\beglet
\begin{eqnarray}
{\bf a^e}& =& ({\bf Y}^T {\bf C}^{-1} {\bf Y})^{-1} {\bf Y}^T{\bf C^{-1}} {\bf x}, \label{M5a} \\
{\bf C} &=& \langle {\bf x} {\bf x}^T \rangle = {\bf S} + {\bf N}, \label{M5b}
\end{eqnarray}
\endlet
(de Oliviera-Costa and Tegmark 2006).  If Gaussianity and statistical
isotropy holds, the variance of the {(\ref{M5a}) is
\begin{equation}
\langle {\bf a}^e {\bf a}^{eT} \rangle = {\bf C_a} = 
({\bf Y^T} {\bf C}^{-1} {\bf Y})^{-1}. \label{M6}
\end{equation}
The statement that the estimator (\ref{M5a}) is `optimal' and the
expression for the variance (\ref{M6}), do of course depend on the
assumptions of Gaussianity and statistical isotropy. However, as
long as the noise term in  (\ref{M3}) is negligible, the reconstructed
harmonic coefficients will be indentical to the true harmonic 
coefficients {\it independent} of any assumptions concerning statistical
isotropy.

\begin{figure*}
\vskip 8.7 truein

\includegraphics{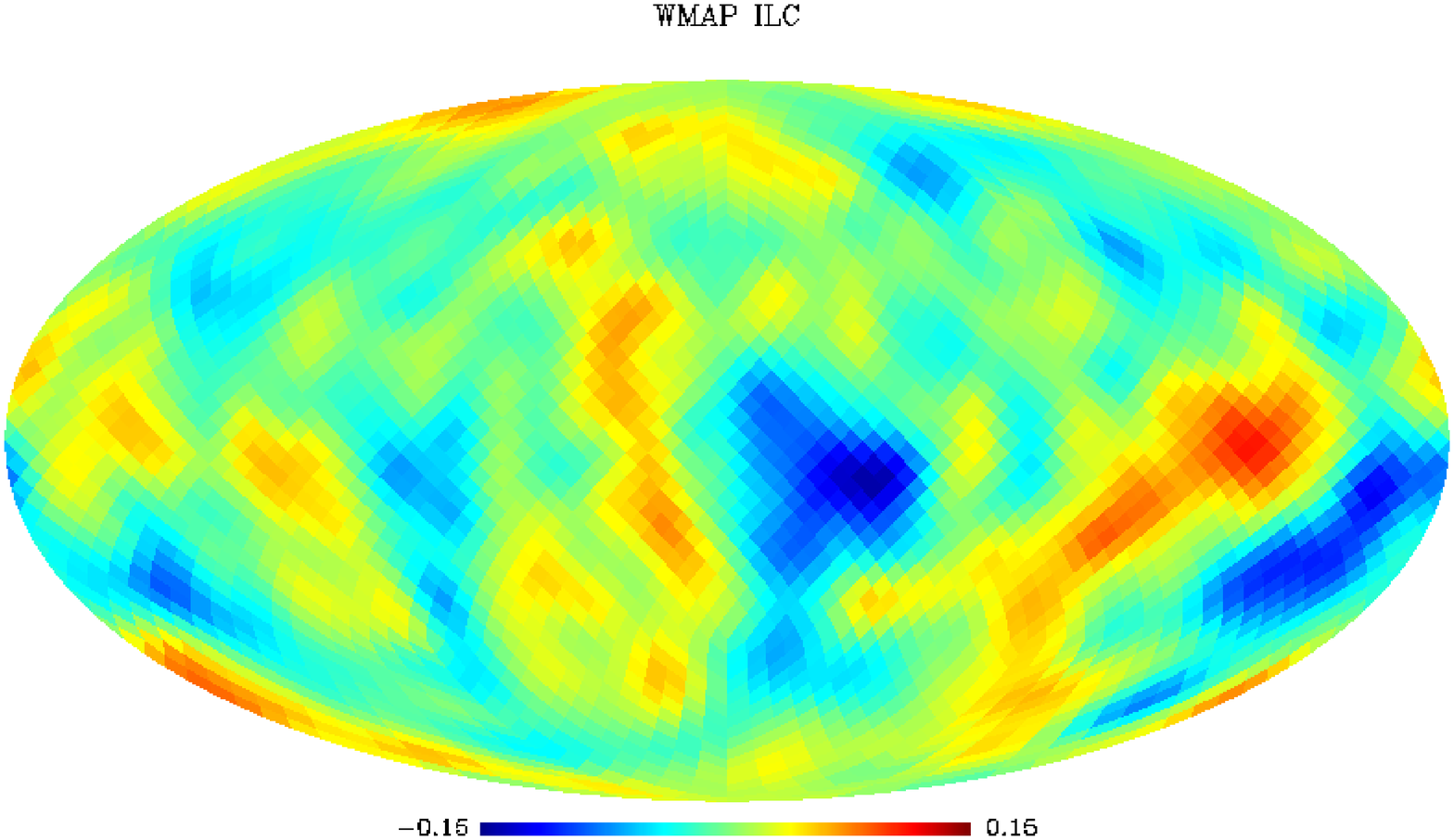}
\includegraphics{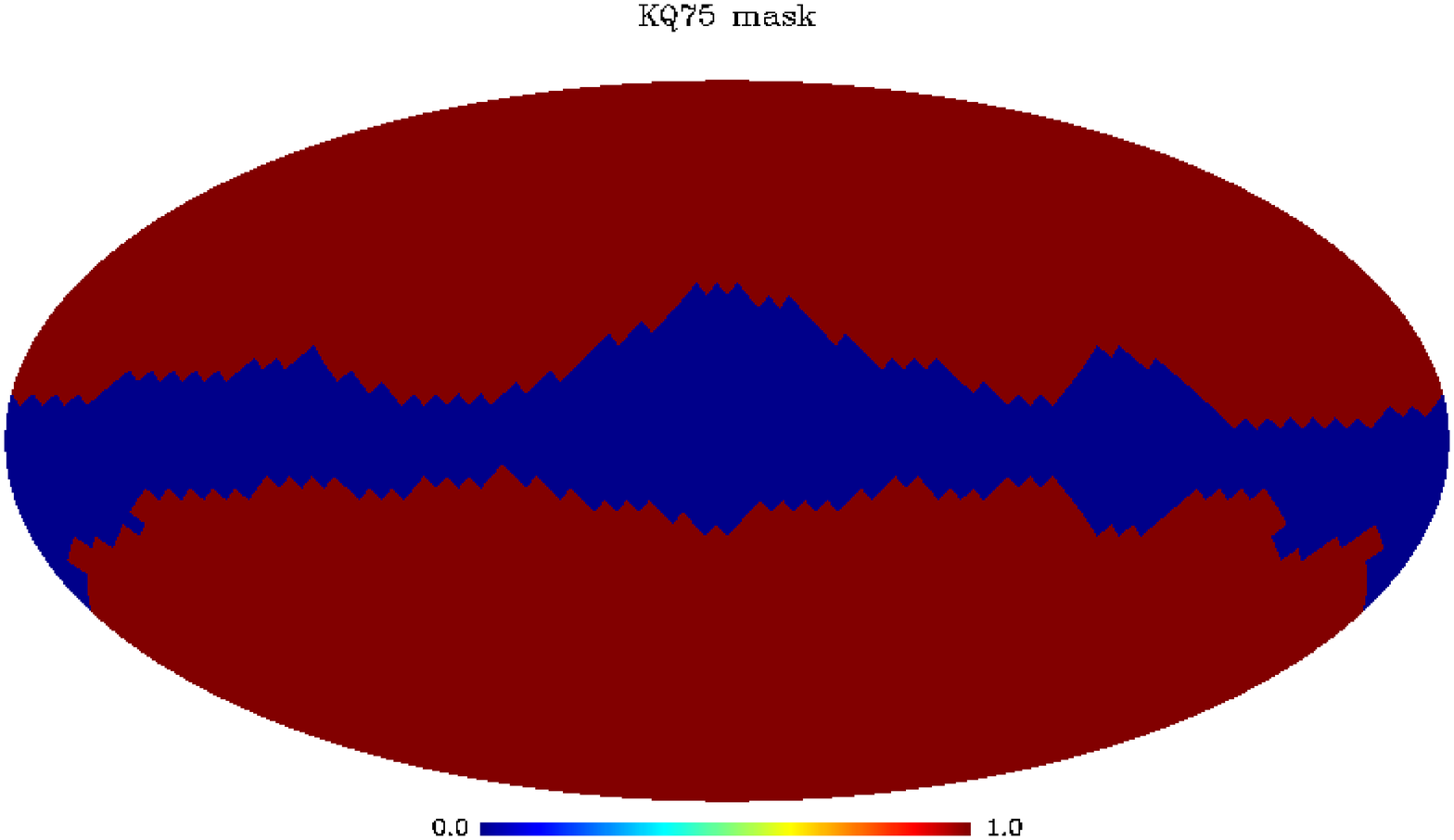}
\includegraphics{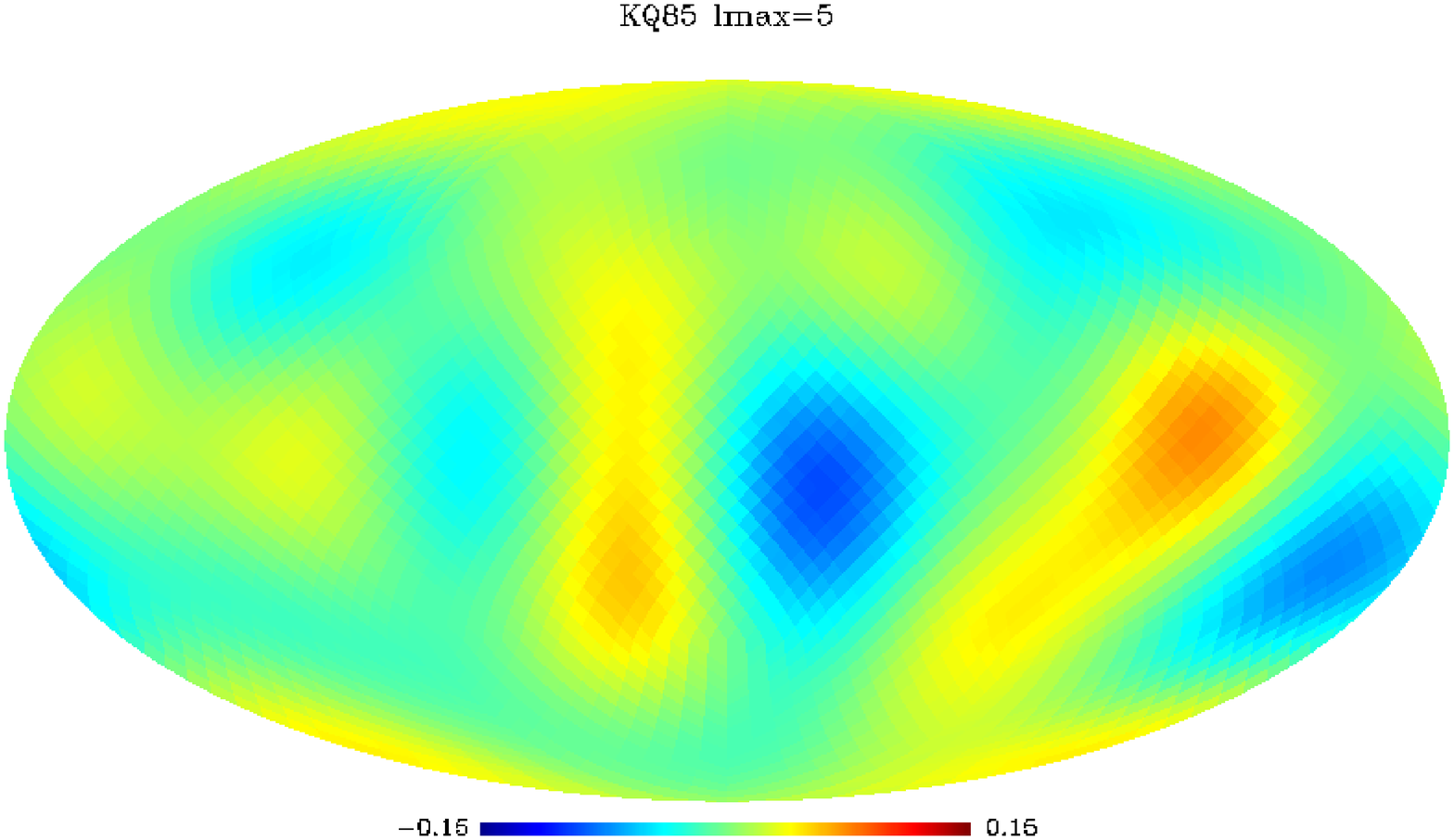}
\includegraphics{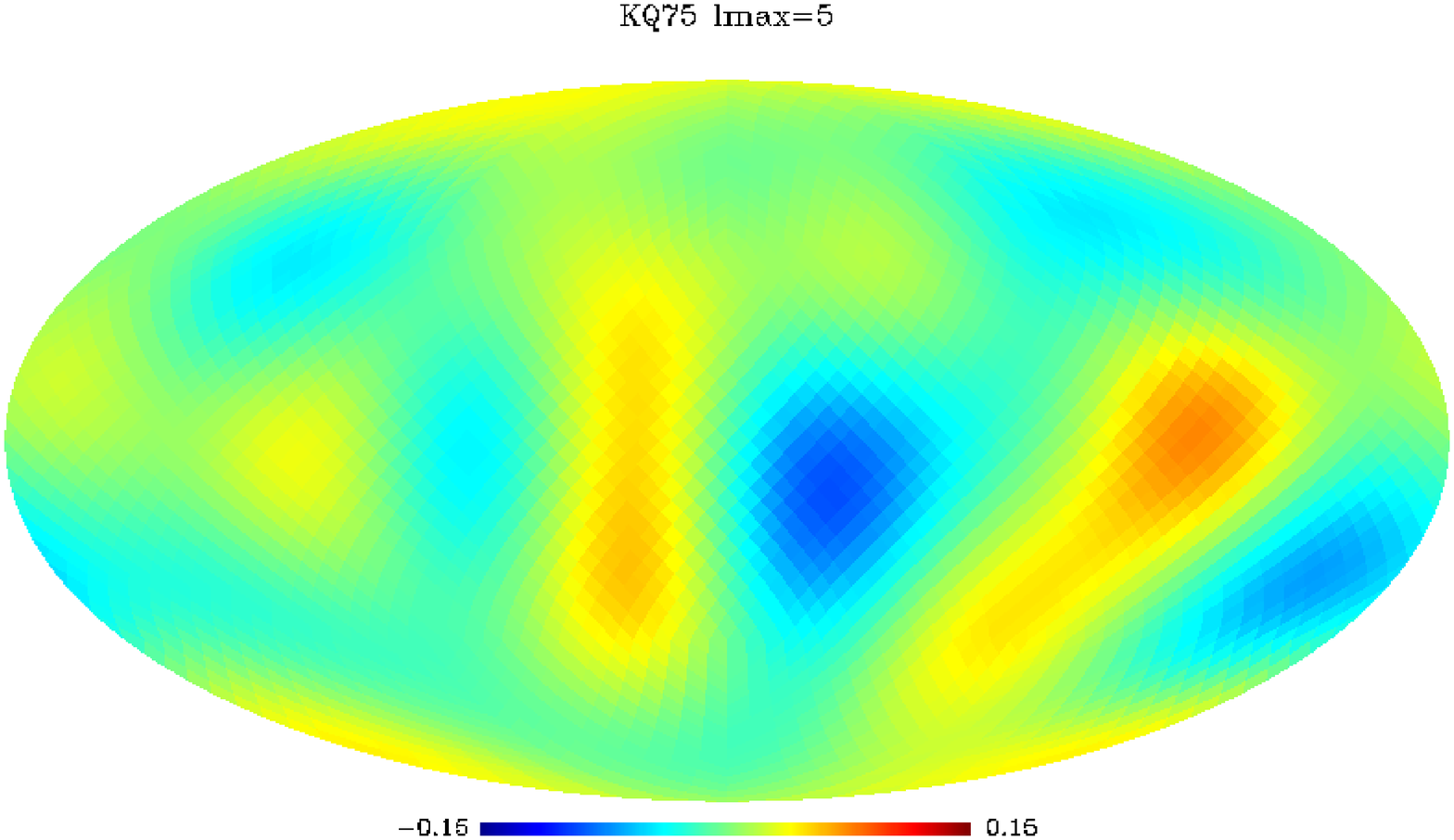}
\includegraphics{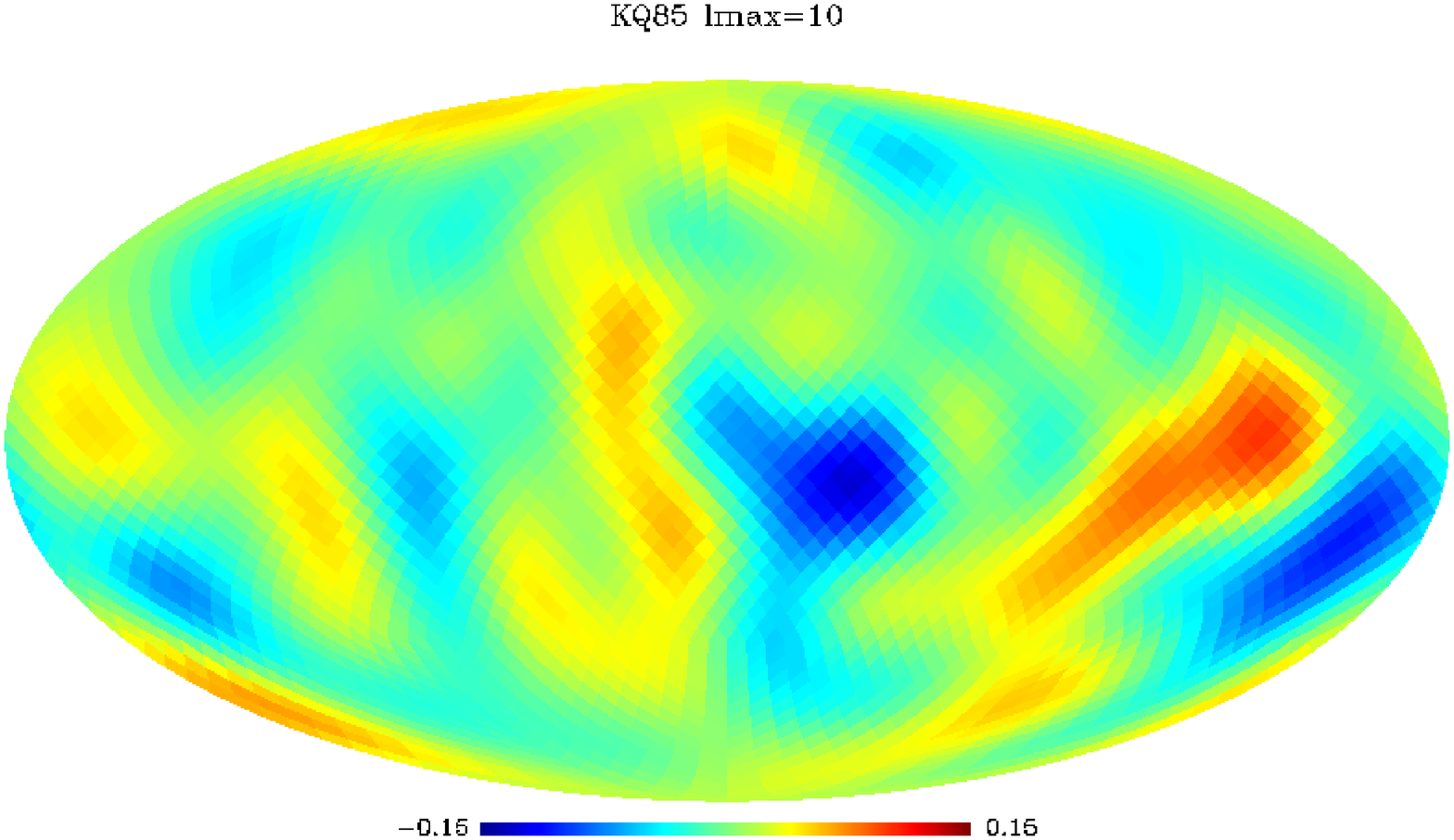}
\includegraphics{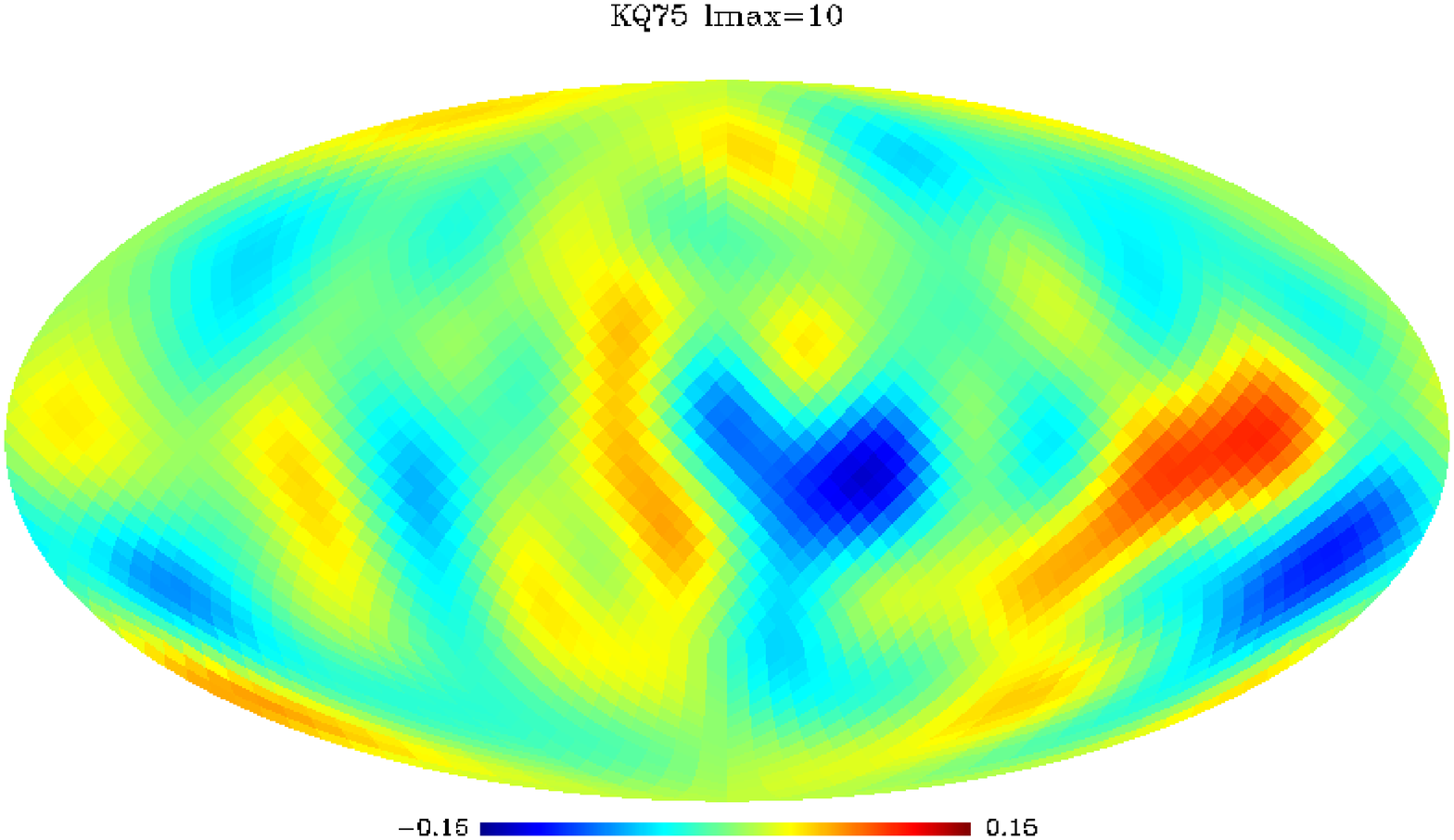}
\includegraphics{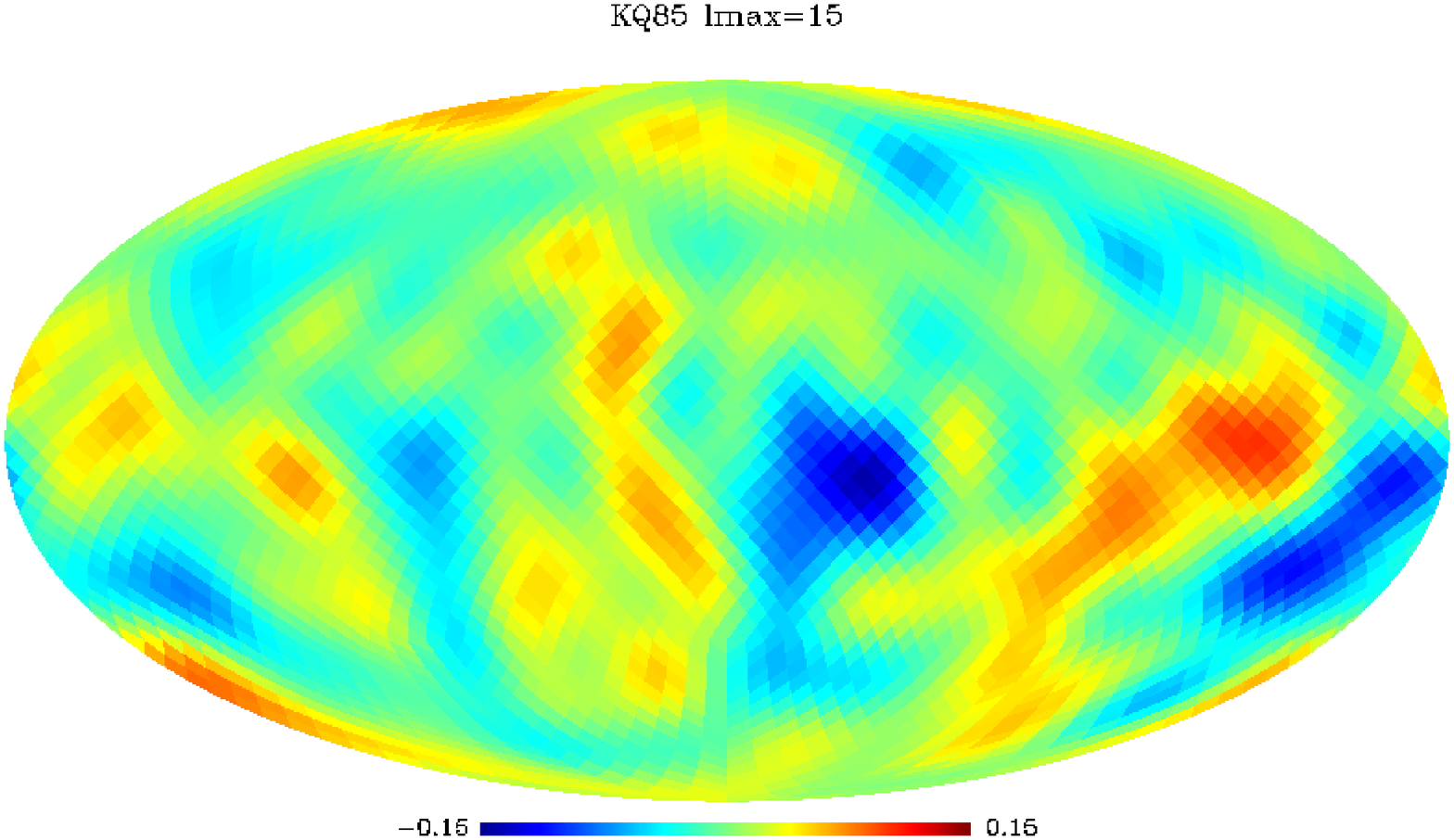}
\includegraphics{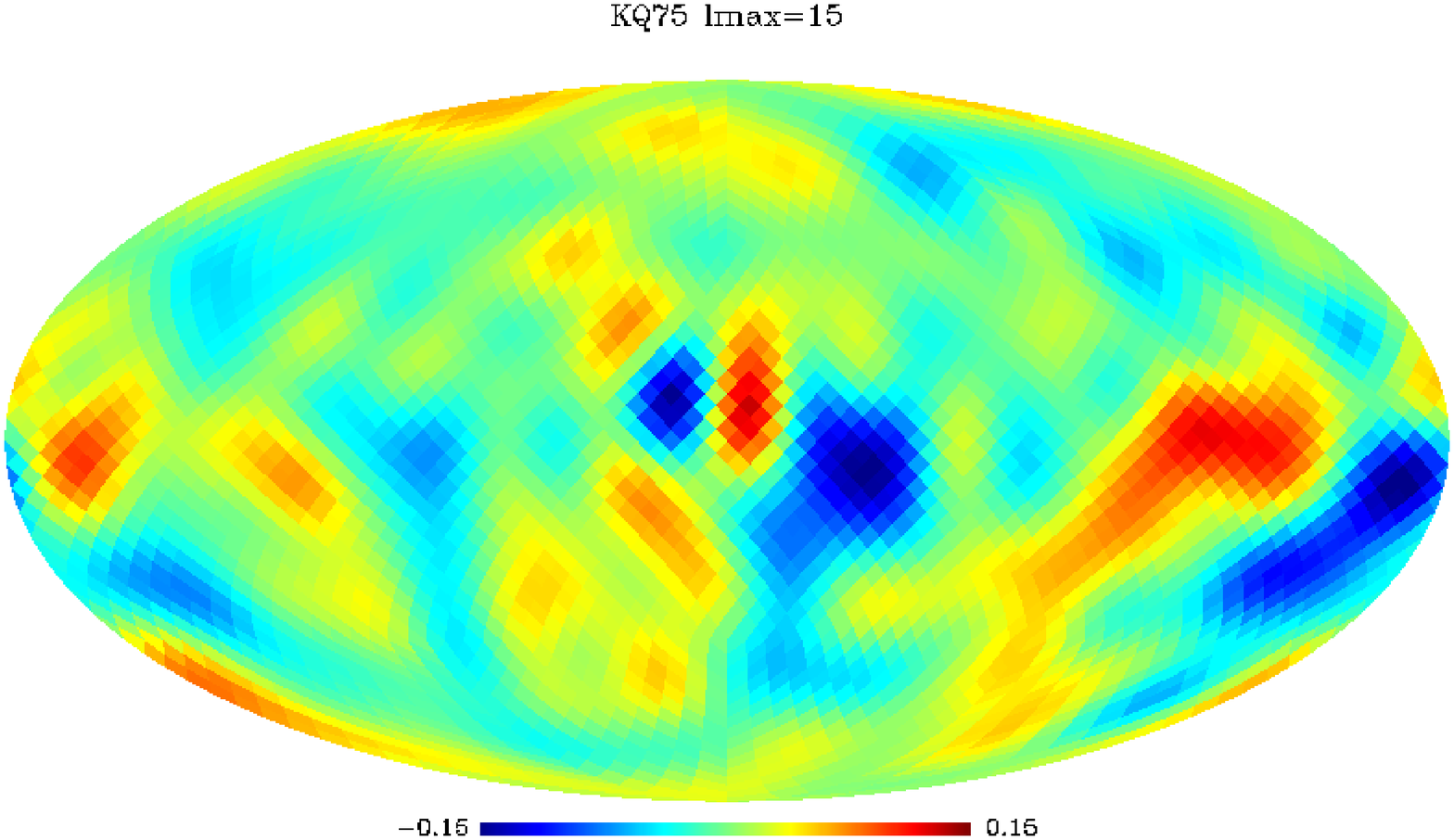}
\includegraphics{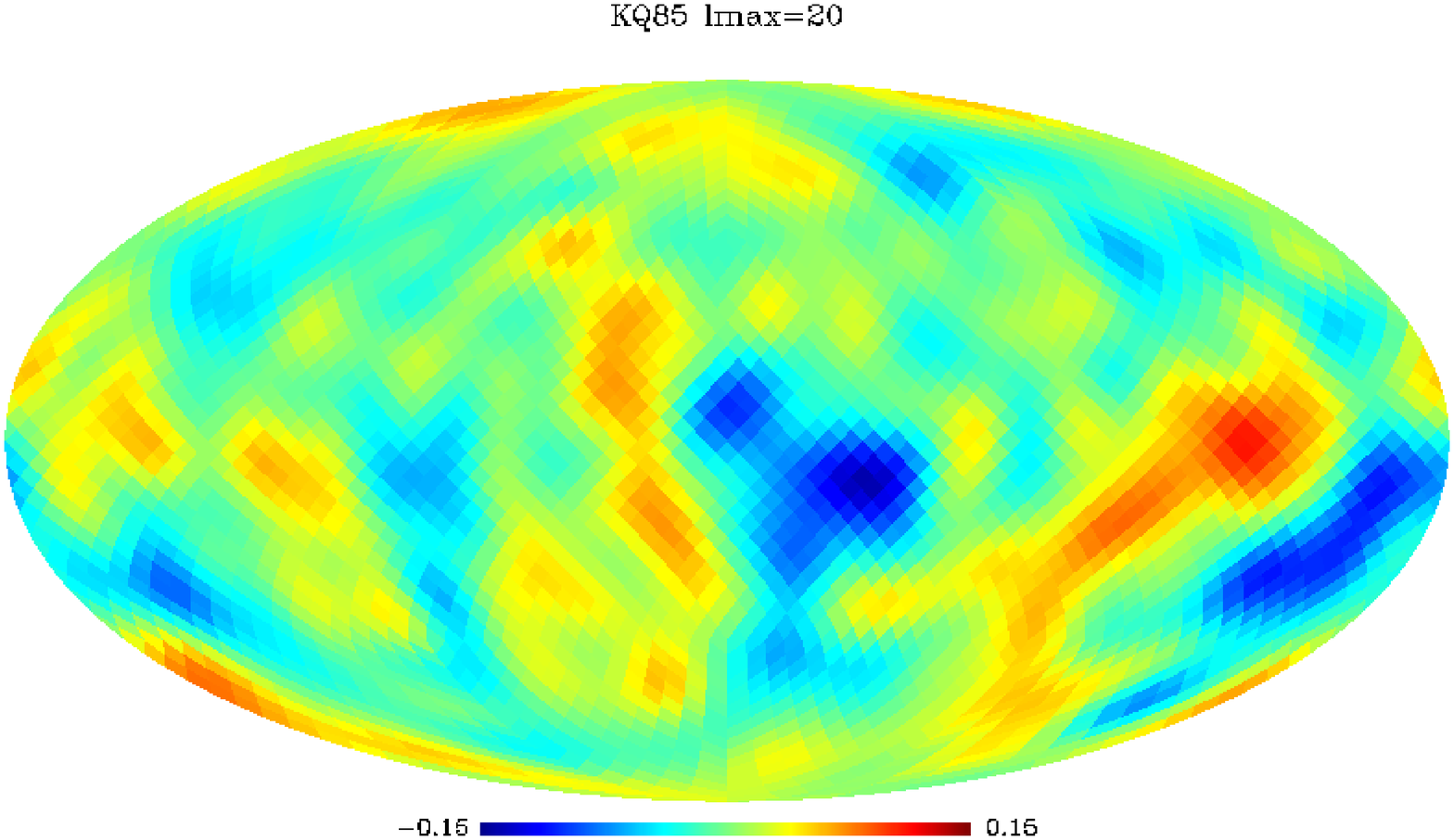}
\includegraphics{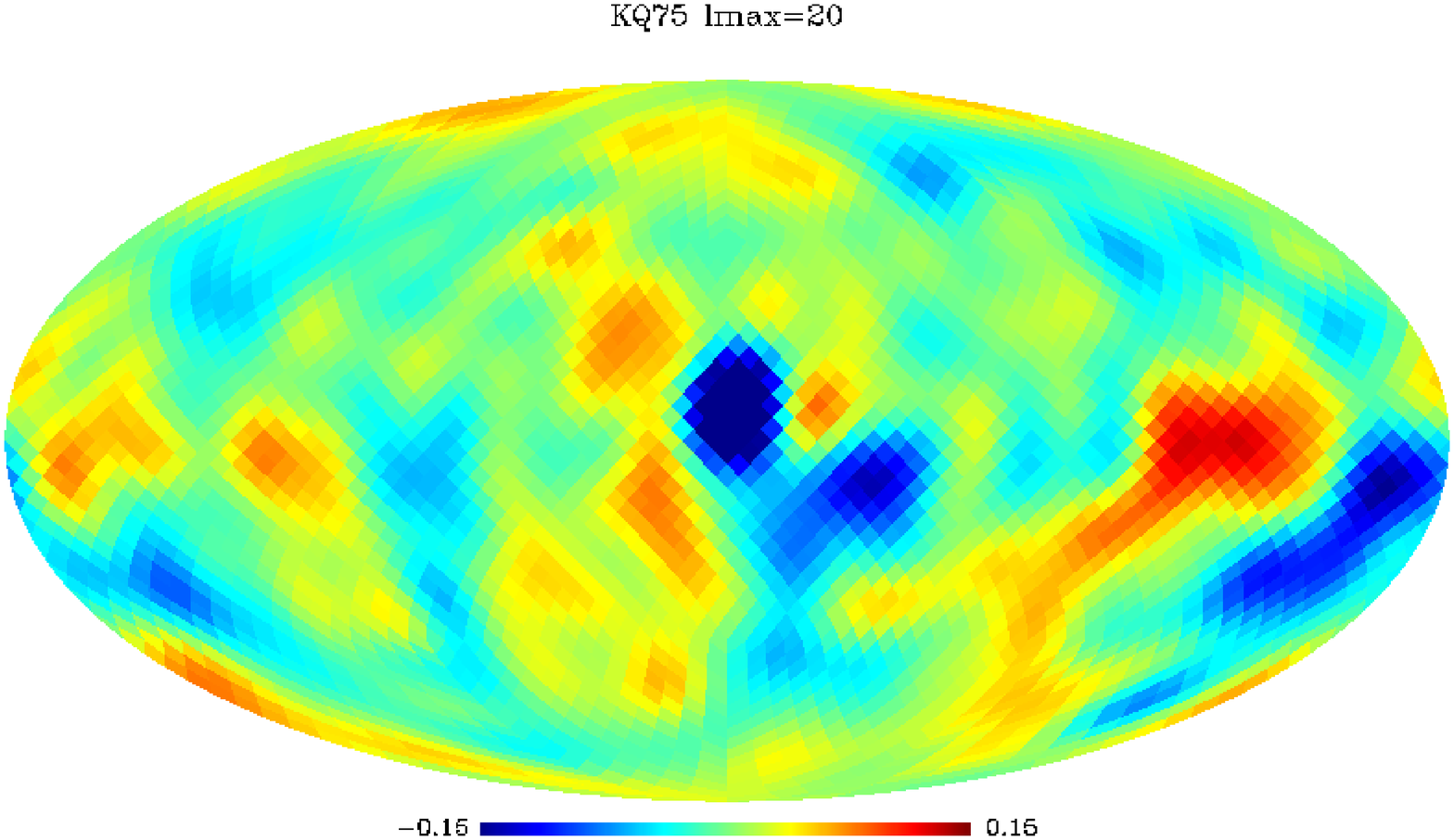}

\caption {The figure to the left in the top row shows the WMAP 5
  year ILC temperature map smoothed by a Gaussian of FWHM $10^\circ$
  and repixelized to a Healpix resolution of NSIDE=16. The figure to
  the right in the top panel shows the degraded resolution WMAP KQ75
  mask used in this paper. The remaining figures to the left show the
  reconstructed all-sky maps computed from data outside the KQ85 
  sky cut,  using the harmonic coefficients computed from equation
  (\ref{M5a}) with the coupling matrices trunctated to (from top to bottom)
   $\ell_{\rm max} = 5$, $10$, $15$ and $20$. The figures to the
  right show equivalent plots for the reconstructed all-sky maps 
  computed from data outside the KQ75 sky cut.}

\label{figure4}
\end{figure*}

Figure 4 illustrates the application of this machinery. The upper
row shows the smoothed WMAP 5 year ILC map (to the left) and the
degraded resolution KQ75 mask (to the right). The remaining figures
show the reconstructed maps from the harmonic coefficients (\ref{M5a})
for the KQ85 mask (figures to the left) and for the KQ75 mask (figures
to the right). The figures show the reconstructions with $\ell_{\rm
  max}$ truncated at $5$, $10$, $15$ and $20$.  The maps for the two
sky cuts at $\ell_{\rm max}=5$ and $10$ are virtually identical, and by
$\ell=10$ the reconstructions look visually similar to the ILC map
over the entire sky. For $\ell_{\rm max}=15$ and $20$, the
reconstructions for the KQ85 mask are stable and, again, look very
similar to the WMAP ILC map over the whole sky. For the KQ75 mask, one
can see `noise' ({\it i.e.} reconstruction errors) beginning to appear
inside the sky cut when $\ell_{\rm max}$ is increased to $\ell_{\rm
  max} =15$ and $20$,

\begin{figure*}
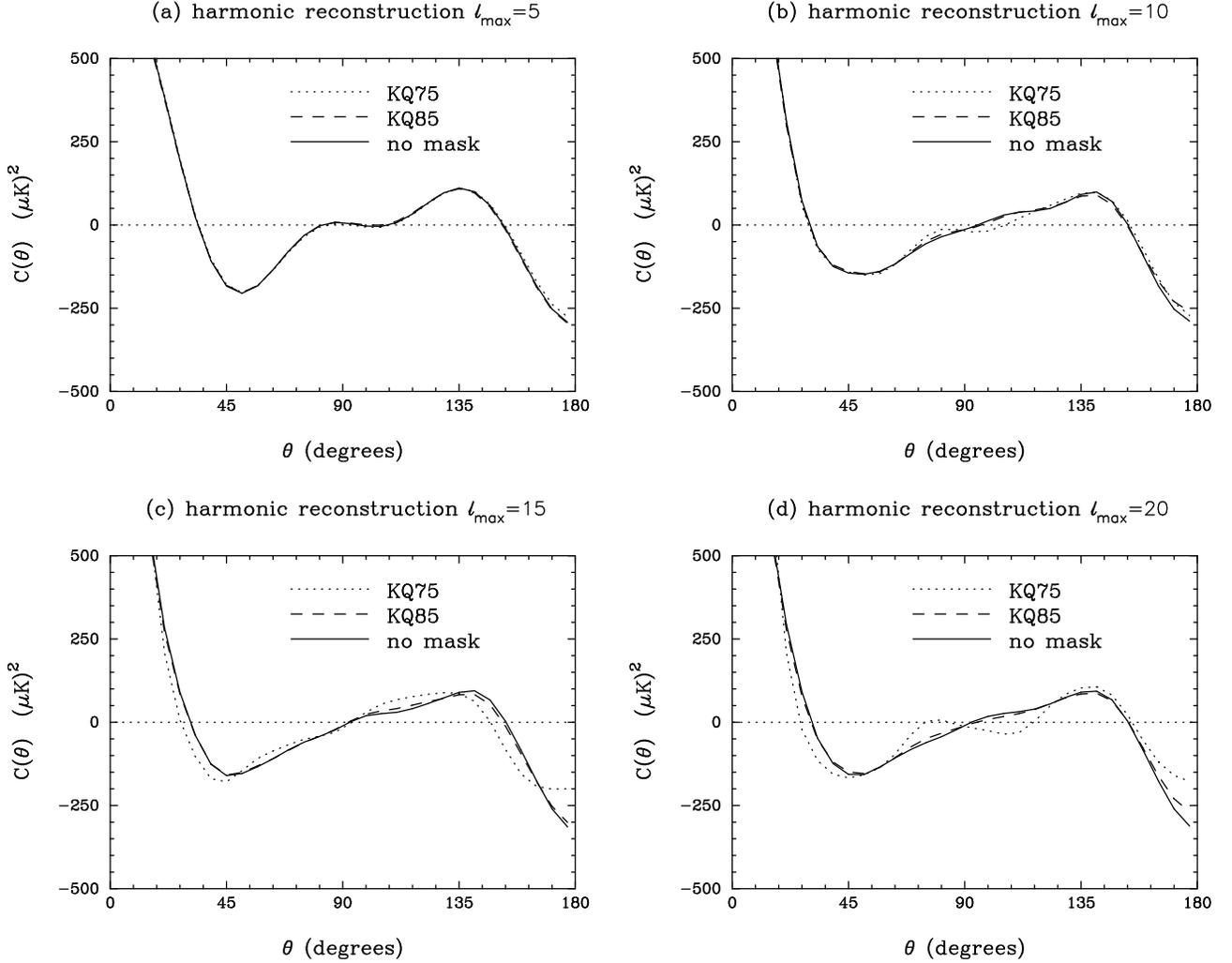

\vskip 5.5 truein

\includegraphics{pgctheta_5cij.ps}
\includegraphics{pgctheta_10cij.ps}

\includegraphics{pgctheta_15cij.ps}
\includegraphics{pgctheta_20cij.ps}

\caption
{Reconstructions of the correlation function from equation (\ref{ML7})
for various choices of $\ell_{\rm max}$ and sky cut.}

\label{figure5}
\end{figure*}

However, the high harmonics that con
tribute to the `noise' in Figure
\ref{figure4} make very little contribution to the correlation
function at large angular scales. This is illustrated in Figure 5,
which shows the dependence of the correlation functions
\begin{equation}
   C^e(\theta)  = {1 \over 4 \pi} \sum_{\ell=2}^{\ell_{\rm max}} (2 \ell + 1)
  C^e_\ell  P_\ell (\cos \theta), \quad 
  C^e_\ell = {1 \over (2 \ell + 1)} \sum_m \vert  a^e_{\ell m} \vert^2,\label{ML7}
\end{equation}
on $\ell_{\rm max}$ for each of the each of the sky cuts.  In the case
of zero sky cut, the correlation function stabilises to its final shape
by $\ell_{\rm max}=10$; higher multipoles make a negligible contribution
to the correlation function at large angular scales. The reconstructed
correlation functions for the KQ85 and KQ75 masks are almost identical
to the all-sky correlation function for $\ell_{\rm max}=5$, $10$ and $15$.
For the KQ75 mask, one can begin to see the effects of reconstruction noise
in $C^e(\theta)$ for $\ell_{\rm max}=20$, but the correlation function for
the KQ85 mask remains stable.

 This analysis shows that it is possible to reconstruct the low
order harmonic coefficients that contribute to the large angle
correlation functions accurately from data on the cut sky. The
sky cut is basically irrelevant and so the all-sky form of the
correlation function can be reconstructed from the cut sky
irrespective of any assumptions concerning Gaussianity or
statistical isotropy. Values for the $S_{1/2}$ statistic
for each of the cases shown in Figure 5 are listed in Table 1.

Notice that if we define weighted harmonic coefficients,
\begin{equation}
\pmb{$\beta$} = {\bf C_a}^{-1} {\bf a^e} = Y^*_{\ell m} (\pmb{$\theta$}_i)
C_{ij}^{-1} x_{j},     \label{QML1}
\end{equation} 
then the power spectrum computed from these weighted coefficients is
\begin{equation}
{y_\ell} = {1 \over 2} \sum_m \vert \beta_{\ell m} \vert^2 = x_p x_q E_{pq}^\ell, \label{QML2}
\end{equation}
where 
\begin{equation}
  {\bf E}^\ell = {1 \over 2} {\bf C}^{-1} {\partial {\bf C} \over \partial C_{\ell}}
{\bf C}^{-1} . \label{QML2a}
\end{equation}
In other words, the power spectrum of the weighted coefficients is identically
equivalent to the QML power spectrum estimator (Tegmark 1997, de Oliviera-Costa
and Tegmark 2006). If statistical isotropy holds, and the data is noise-free,
the quantity
\begin{equation}
{\bf \hat C^Q} = {\bf F}^{-1} {\bf y}, \label{QML3}
\end{equation}
provides an unbiased estimate of the power spectrum, where ${\bf F}$ is the
Fisher matrix,
\begin{equation}
F_{\ell \ell^\prime} = {1 \over 2} {\rm Tr} \left [{\bf C^{-1}} {\partial {\bf C} \over \partial C_{\ell}}
{\bf C^{-1}} {\partial {\bf C} \over \partial C_{\ell^\prime}} \right ]. \label{QML4}
\end{equation}
Notice that for the complete sky, and for noise-free data
\begin{equation}
{\bf C_a} = C_\ell \delta_{\ell \ell^\prime} \delta_{m m^\prime},  \label{QML5}
\end{equation}
in the limit $\ell_{max} \rightarrow \infty$, {\it i.e.} the variance on the $a_{\ell m}$
is just the cosmic variance. The Fisher matrix is 
\begin{equation}
\quad F_{\ell \ell^\prime}
= {(2 \ell + 1) \over 2 C_\ell^2} \delta_{\ell \ell^\prime}, \label{QML6}
\end{equation}
and so the QML estimates $\hat C^Q_{\ell}$ are identical to the PCL
estimates.  For relatively small sky cuts such as the KQ85 and KQ75
masks, the Fisher matrix at low multipoles will be {\it almost}
diagonal (see Efstathiou 2004b) and the recovered power spectrum from
the cut sky will be almost identical to the true power spectrum
computed from the whole sky. The QML estimator effectively performs
the reconstruction ${\bf a^e}$ of equation (\ref{M5a}), but uses the
assumption of statistical isotropy to downweight `ambiguous' modes
that are poorly constrained by the sky cut.

For small sky cuts, we would therefore expect the QML correlation
function estimate (\ref{C13}) to be almost identical at large-angular
scales to the correlation functions computed from the reconstructed
coefficients ${\bf a^e}$. (They are, of course, mathematically
identical for zero sky cut.) $C^Q(\theta)$ is expected to behave more
stably than $C^e(\theta)$ as the sky cut is increased, since the QML
correlation function downweights ambiguous modes.  This is exactly
what we see when we apply the QML estimate to the WMAP ILC 5 year ILC
maps (see Figure \ref{figure6}). The angular correlation function is almost
independent of the sky cut, confirming the results of E04. Values
of the $S_{1/2}$ statistic for the QML ACF estimates are listed in the
final column of Table 1.

\begin{figure*}
\vskip 3.0 truein

\includegraphics{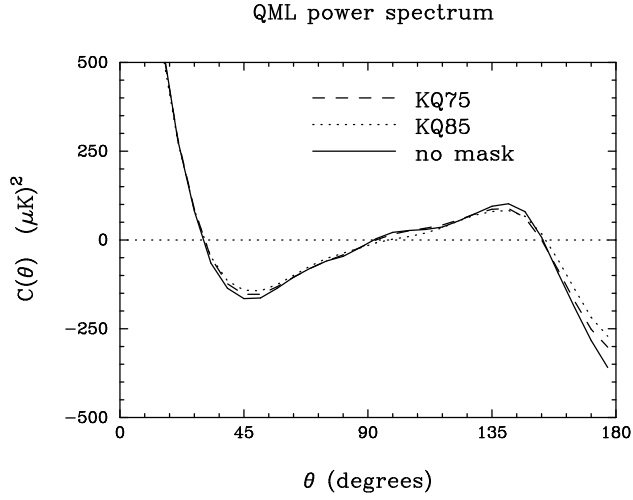}

\caption
{As Figure 1 but using the QML estimator of equation (\ref{C13}).}

\label{figure6}
\end{figure*}

\begin{figure*}
\vskip 3.0 truein

\includegraphics{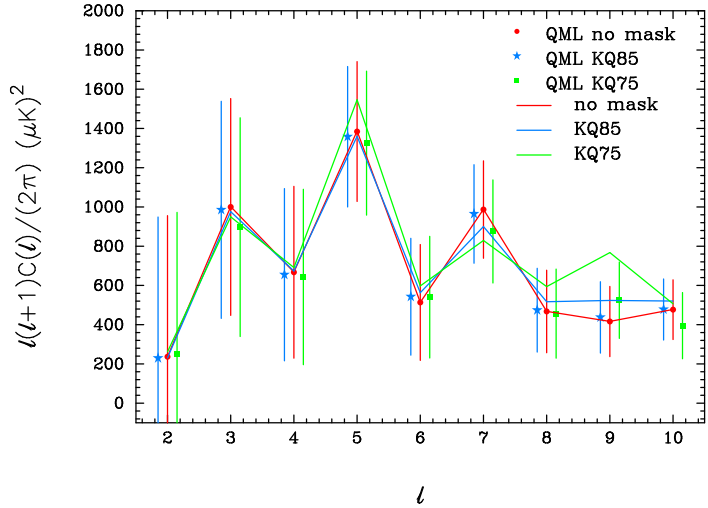}

\caption {The temperature power spectrum at low multipoles computed
  from the low resolution WMAP 5 year ILC map. The points (corrected
  for the $10^\circ$ FWHM smoothing and slightly displaced in $\ell$
  for clarity) show QML power spectrum estimates for three sky cuts:
  no mask; KQ85 mask; KQ75 mask.  Error bars show the diagonal
  components of the inverse of the Fisher matrix (\ref{QML4}). The
  solid red line shows the power spectrum computed from the all-sky
  ILC map (which is identical to the QML all-sky estimates). The solid
  green and blue lines show the power spectra computed for the
  $\ell_{\rm max} = 10$ reconstructions of Figure 4.}

\label{figure7}
\end{figure*}

The QML power spectrum estimates are plotted in Figure \ref{figure7}. The power 
spectrum coefficients $\hat C^Q_\ell$ are extremely stable to the sky cut,
varying by only a few tens of $(\mu{\rm K})^2$ for $\ell \le 10$. The figure
compares these estimates to the power spectrum estimates for the reconstructed
all-sky maps using equation (\ref{M5a}). We plot the results for $\ell_{\rm max}=10$,
since this value is large enough to determine the shape of the ACF at large
angular scales, but small enough to limit the noise in the reconstructed
maps at high multipoles. The power spectra of the reconstructed maps are
very close to the QML estimates at $\ell \le 8$, though one can begin to see the
effects of reconstruction noise in the KQ75 case at $\ell > 8$. (However,
as Figure \ref{figure5}b shows, this reconstruction noise has very little
effect on the shape of the ACF at large angular scales.)

The results of this section show that the low-order multipole
coefficients that determine the behaviour of the correlation function
at large angular scales can be reconstructed to high accuracy from data on the
incomplete sky, independent of any assumptions concerning statistical
isotropy. The usual motivation for applying a sky cut is to remove
regions of the sky that may be contaminated by residual Galactic
emission. However, for the KQ85 and KQ75 sky cuts, the missing area of
sky leads to little loss of information at low multipoles. The low
multipoles can therefore be reconstructed  from the
data on the incomplete sky. The imposition of the sky cuts does not
remove foreground contamination at these low multipoles:
any residual Galactic contribution to the low multipoles in the ILC
map is, like the CMB signal, faithfully reproduced by the
reconstructions shown in Figure \ref{figure4}.  What a Galactic cut
can do is to mask out localised Galactic emission (`ambigous' modes)
that could, in principal,  couple to the low multipoles in a way that depends on the
estimator ({\it e.g.} via the coupling matrix ${\bf K}$ in the simple
inversion of equation (\ref{M4})).  The similarities between the
reconstructions of Figure 4 and the full-sky ILC map (and the
correlation functions and power spectra plotted in Figures 5-7) show
that the ILC map has removed Galactic emission successfully at low
Galactic latitudes, since there is no evidence of high amplitude
`ambiguous' modes in the ILC map within the region of the sky cut.

If suitable estimators are applied to noise-free data, a sky cut of
the size of the KQ85 or KQ75 masks has little impact on the
reconstruction of the low order multipoles or the all-sky ACF.  The
imposition of a sky cut does, however, lead to a loss of information
if a poor estimator is used to estimate the ACF. This is what happens
when the pixel estimator (\ref{C1}) is used to estimate the ACF on the
cut sky (Copi \etals 2007; Hajian 2007; CHSS09).  The analysis
presented in this Section provides compelling evidence that the true
value of the $S_{1/2}$ statistic for our realization of the sky is in
the region of $6000$-$8000 (\mu {\rm K})^4$, independent of the sky
cut.  

\section{Analysis of the S$_{\bf 1/2}$ statistic}

In this Section we analyse the  $S_{1/2}$ statistic, first
from a Bayesian point of view, and then from a frequentist
point of view. We then discuss the interpretation of the low frequentist
p-values found by CHSS09.

\subsection{Approximate Bayesian analysis}

We begin by  performing an approximate Bayesian analysis to
compute the posterior distribution of the $S_{1/2}$  given
the data on the assumption that the fluctuations are Gaussian and
statistically isotropic. If the data were noise-free and covered the
entire sky then, under the assumptions of statistical isotropy and
Gaussianity, the data power spectrum $C^d_\ell$ provides a loss-free
description of the data. Assuming uniform priors on each of the
$C_\ell^T$, the posterior distribution of the theory power spectrum
coefficients $C^T_{\ell}$ is given by the inverse Gamma distribution
\begin{equation}
dP(C^T_\ell\vert C^d_\ell) \propto \left (C^d_\ell \over C^T_\ell \right )^{2 \ell - 1 \over 2}
{\rm exp} \left [ -{(2 \ell + 1)  \over 2} \left ({ C^d_\ell  \over C_\ell^T} \right ) \right ] {1 \over C_\ell^T}. \label{B1}
\end{equation}
Each of the $C^T_\ell$ is statistically
independent and  the mean value is
\begin{equation}
\langle C^T_\ell \rangle  = \left ( {2 \ell +1 \over 2 \ell - 3} \right ) C^d_\ell. \label{B2}
\end{equation}
The distribution (\ref{B1}) will therefore favour theory values that are 
larger than the observed values $C^d_\ell$.

The results of the previous Section (and Figure \ref{figure7} in
particular) show that the low multipoles are well determined and
insensitive to the application of a sky cut.  We can
therefore use the measurements $C^d_\ell$ computed over the whole sky
to represent the data\footnote{It is in this sense that the analysis
  presented here is described as an `approximate', {\it i.e.} any
  residual errors on the $C^d_\ell$ are ignored.}.  The multipole
expansion is truncated at $\ell_{\rm max}=20$ (although as discussed
in the previous Section, multipoles greater than $\ell \approx 10$
make very little contribution to the ACF at large angular scales) and
statistically independent $C^T_\ell$ values are generated from the
inverse Gamma distribution ({\ref{B1}).  These values are then used to
  generate Gaussian $a^T_{\ell m}$ from which we synthesize real-space
  maps $x_i$ at a Healpix resolution of NSIDE=16 smoothed with a
  Gaussian of FWHM $10^\circ$. We then compute $S_{1/2}$ from the
  pixel correlation function (\ref{C1}).  This methodology provides
a test of statistically isotropic, Gaussian models, with no additional
constraints imposed on the theory $C^T_{\ell}$ apart from 
uniform priors. 

The posterior distributions of $S^T_{1/2}$ is shown in Figure
\ref{figure8}a for the analysis of all-sky maps (red histogram) and
for maps with the KQ75 mask applied (blue histogram). The distribution
for the trials with the sky cut applied is slightly broader than the
distribution for the all-sky trials, as expected since the pixel ACF
estimator is sub-optimal on a cut sky.  The peaks of the distributions
occur at $S^T_{1/2} \approx 6000 \;(\mu {\rm K})^4$ and so low values
of $S^T_{1/2}$ are clearly preferred by the data. However, the
posterior distributions have a very long tail to high values (as
expected from the inverse Gamma distribution \ref{B1}). The best
fitting six parameter $\Lambda$CDM model as determined from the 5 year
WMAP analysis (Komatsu \etals 2009) has a value of $S^T_{1/2} \sim
49000 \; (\mu {\rm K})^4$. At this value (indicated by the vertical
dashed line in Figure \ref{figure8}a), the posterior distribution has
fallen to a value of about $0.4$. Such high values of $S^T_{1/2}$ are
evidently not favoured by the data, but they are not strongly
disfavoured. Very low values of $S^T_{1/2}$, of $\sim 1000 \; (\mu
{\rm K})^4$ are also not strongly disfavoured.

  From the Bayesian point of view, the quantity $S_{1/2}$ is a poor
  discriminator of theoretical models and so is relatively
  uninformative. The posterior distributions of Figure \ref{figure8}a
  are extremely broad with a long tail to high values. The data,
  irrespective of estimator or sky cut, clearly prefer low values of
  $S_{1/2}$ but cannot exclude the value of $S^T_{1/2} \sim 49000 \;
  (\mu {\rm K})^4$ expected for the concordance inflationary
  $\Lambda$CDM model.

\begin{figure*}
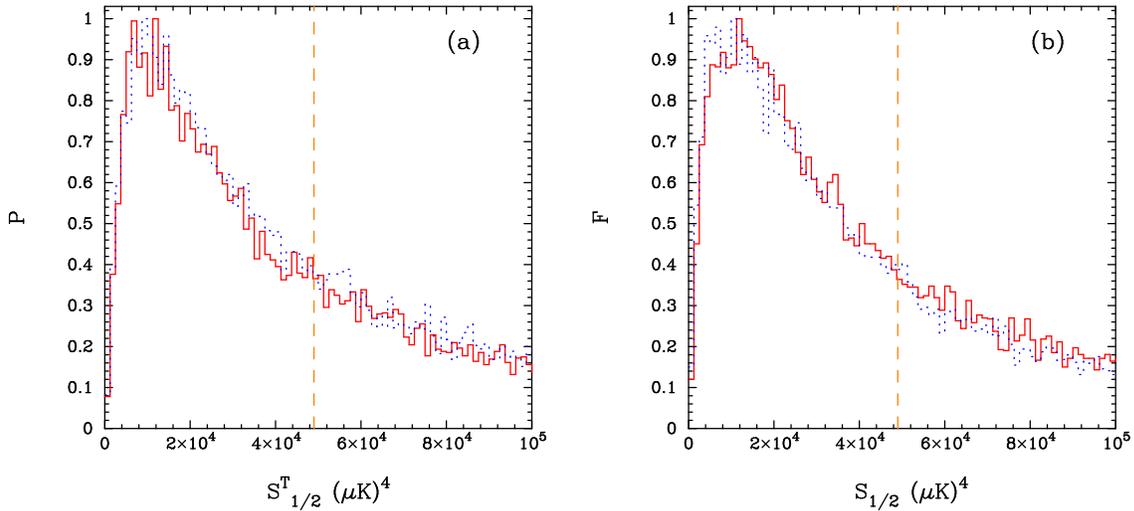

\vskip 3.0 truein

\includegraphics{pgbayes.ps}
\includegraphics{pgfreq.ps}

\caption {(a) Posterior distributions of $S^T_{1/2}$ computed as discussed
  in the text.  The red (solid) histogram shows the distribution of
  $S^T_{1/2}$ from an analysis of the whole sky. The blue (dotted) hisogram
  shows the distribution computed with the KQ75 sky cut applied. The
  vertical dashed lines in the figures shows the value $S^T_{1/2} \sim 49000 \; (\mu
  {\rm K})^4$ for the best fitting $\Lambda$CDM model as determined
  from the 5 year WMAP analysis (Komatsu \etals 2009).
(b) Frequency distributions of $S_{1/2}$ for statistically
  isotropic, Gaussian, realizations of the (Komatsu \etals 2009)
  $\Lambda$CDM model.  The red (solid) histogram shows the frequency
  distribution for the pixel ACF estimator applied to the whole
  sky. The blue (dotted) hisogram shows the distribution computed with
  the pixel ACF estimator with the KQ75 sky cut applied. }

\label{figure8}
\end{figure*}

\subsection{Frequentist analysis}

We now generate statistically isotropic Gaussian realizations with the
$C^T_\ell$ constrained to those of the best fitting $\Lambda$CDM
model. The frequency distributions of $S_{1/2}$ computed from the
pixel estimator are plotted in Figure \ref{figure8}b. The
distributions of Figures \ref{figure8}a and \ref{figure8}b look fairly
similar, but the frequentist interpretation is very different. For the
all-sky analysis, the p-value of finding $S_{1/2} < 7373 \; (\mu {\rm
  K})^4$ is $8\%$ and hence is not statistically significant. However,
if we apply the KQ75 mask, the p-value for $S_{1/2} < 647 \; (\mu {\rm
  K})^4$ is only $0.065\%$.  This result appears strongly significant
and, at face value, inconsistent with the p-value for the all-sky
analysis.

The low p-value found here and by S03 and CHSS09 come exclusively from
analysing cut sky maps with `sub-optimal' (in the sense of not
reproducing the ACF for the whole sky) estimators. The sky cuts,
ostensibly imposed to reduce any effects of Galactic emission at low
Galactic latitudes, lead to a loss of information and to poorer
estimates of the ACF for our realization of the sky. But as we have
demonstrated, the information on the ACF at large angles for our
realization of the sky is contained in the data outside the sky
cuts. The imposition of a sky cut therefore has nothing to do with
reducing the effects of Galactic emission on the ACF at large angular
scales.  If there is any cosmological significance to the low
p-values, then one must accept that the Galactic cut aligns with the
signal, {\it purely by coincidence}, in just such a way as to remove
the large-scale angular correlations for particular choices of
estimator of the ACF. This alignment may indicate a violation of
statistical isotropy, as argued by CHSS09, but if this is true
the alignment with the Galactic plane must be purely coincidental.

It seems to us that a more plausible interpretation of the low
p-values is that they are a consequence of the {\it a posteriori}
selection of the $S_{1/2}$ statistic by Spergel \etals (2003) for a
particular choice of estimator and sky cut. It is difficult to
quantify the effects of {\it a posteriori} choices. However, numerical
tests with the more general statistic
\begin{equation}
   S^p_\mu = 
\left [ {2 \over 3(1+\mu)} \int_{-1}^\mu [C(\theta)]^p \; d\cos \theta \right ]^{2/p}, \label{CC1} 
\end{equation}
(which reduces to $S_{1/2}$ for the choices $\mu=1/2$ and $p=2$)
suggest that it is possible to alter p-values by an order of magnitude
or more by selecting the parameters in response to the data. It would
be possible to raise the p-values even more by varying the size and
orientation of a sky cut.

\begin{figure*}
\vskip 3.0 truein

\includegraphics{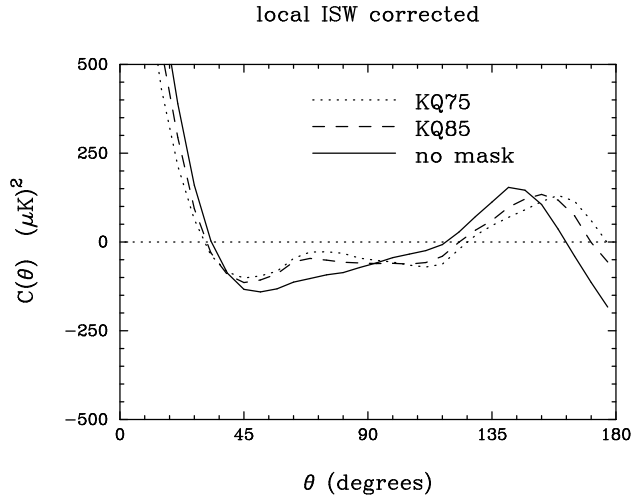}

\caption {Pixel based estimates ACF estimates for the WMAP ICL map
corrected for the local ISW contribution for redshift $z<0.3$
as described by Francis and Peacock (2009). }

\label{figure9}
\end{figure*}

Is there any way of testing this hypothesis further ? In the $\Lambda$CDM model,
the integrated Sachs-Wolfe (ISW) effect (Sachs and Wolfe, 1967) makes
a significant contribution to the total temperature anisotropy signal  
at low multipoles. The ISW contribution from the time of last scattering
($t_{\rm LS}$) and the present day ($t_0$)  is given by
\begin{equation}
  {\Delta T \over T}^{\rm ISW} = 2 \int_{t_{\rm LS}}^{t_0} {d\Phi \over dt} \ dt
\end{equation}
where $\Phi$ is the Newtonian gravitational potential (see {\it e.g.}
Muhkanov 2005). Recently, Francis and Peacock (2009) have used the
2MASS near infrared all-sky survey (Jarrett 2004), together with
photometric redshift estimates to compute the ISW contribution from
local structure at redshifts $z<0.3$.  If {\it a posteriori} choices
are responsible for the low p-values, we should find large changes to
the pixel ACF estimates for the masked sky when the WMAP ILC maps are
corrected for the local ISW contribution.  As shown in Figure 9, this
is indeed what we find when we subtract the local ISW contribution
computed by Francis and Peacock from the 5 year WMAP ILC map. The
$S_{1/2}$ statistic computed from the ACFs shown in Figure 9 are
$10360 \; (\mu {\rm K})^4$ (all-sky), $6463 \; (\mu {\rm K})^4$ (KQ85
mask) and $5257 \; (\mu {\rm K})^4$ (KQ75 mask), all consistent with
the concordance $\Lambda$CDM model at the few percent level.  This is
consistent with our hypothesis that the CHSS09 low p-values are a fluke,
unless one is prepared to argue that there is a physical alignment of
local structure with the potential fluctuations at the last scattering
surface that conspires to remove large-angle temperature correlations
in the regions outside the Galactic mask (which seems
implausible). Francis and Peacock (2009) discuss how the local ISW
correction affects a number of other low multipole statistics, in
particular reducing the statistical significance of the alignment
between the quadrupole and octopole.

\section{Discussion and Conclusions}

The low amplitude of the temperature autocorrelation function at large
angular scales has led to some controversy since the publication of
the first year results from WMAP. This paper has sought to
clarify the following points:

\smallskip

\noindent
[1] We have compared different estimators of the ACF showing: (a) how
they depend on assumptions of statistical isotropy; (b) how they are
interrelated; (c) how they perform on the WMAP 5 year ILC maps with
and without a sky cut.

\smallskip

\noindent 
[2] The imposition of the KQ85 and KQ75 sky masks leads to little loss
of information on the low multipoles that contribute to the
large-scale angular correlation function. As demonstrated in Section
3, the low multipole harmonics can be reconstructed accurately,
independent of any assumptions concerning statistical isotropy, from
the data that lie outside the sky cuts. The ACFs computed from these
reconstructions are in good agreement with the ACF computed from the
whole sky and in good agreement with the maximum likelihood estimator
(\ref{C13}). There can be little doubt that the large-scale ACF for
our realization of the sky is very close to the all-sky results shown
in Figures 1, 3 and 6, independent of any assumptions regarding
statistical isotropy.

\smallskip

\noindent [3] The Bayesian analysis presented in Section 4 shows that
the posterior distribution of the $S^T_{1/2}$ is broad and cannot
exclude the value $S^T_{1/2} \sim 49000 \; (\mu {\rm K})^4$
appropriate for the Komatsu \etals (2009) best fitting inflationary
$\Lambda$CDM model.  The breadth of the posterior distribution
$S^T_{1/2}$ distribution shows that it is fairly uninformative
statistic and so is not a particularly good discriminator of
theoretical models.

\smallskip

\noindent [4] Unusually low values of the $S_{1/2}$ statistic are
found only if `sub-optimal' ACF estimators are applied to maps that
include a Galactic mask. We have argued that the low $p$-values
associated with these low values of $S_{1/2}$ are most plausibly a
result of {\it a posteriori} choices of statistic. This seems
plausible because: (a)  $S$-type statistics are relatively
uninformative and hence sensitive to {\it a posteriori} choices; (b)
the all-sky ACF (which is compatible with the concordance $\Lambda$CDM
model) can be recovered from the data outside the mask and so any
physical model for the low $p$-values requires a fortuitous alignment
of the temperature field with the Galaxy; (b) the analysis of the
local ISW corrected maps presented in Section 4 suggests that any
physical model of the low $p$-values requires a precise alignment of
local structure with the large-scale potential fluctuations at the
last scattering surface.

\medskip

In summary, the results of this paper suggest that, irrespective of
the imposition of Galactic sky cuts or assumptions of statistical
isotropy, the large-scale correlations of the CMB temperature field
provide unconvincing evidence against the concordance inflationary
$\Lambda$CDM cosmology.

\vskip 0.1 truein

\noindent {\bf Acknowledgments:} 
The authors acknowledge use of the Healpix package and the Legacy
Archive for Microwave Background Data Analysis (LAMBDA). Support for
LAMBDA is provided by the NASA Office of Space Science. We are
particular grateful to John Peacock and Caroline Francis for allowing
us to use their ISW maps.  Yin-Zhe Ma thanks Trinity College Cambridge
and Cambridge Overseas Trusts for support. Duncan Hanson is grateful
for the support of a Gates scholarship.

\end{document}